\documentclass{ws-ijmpc}
\def\wsdraftnote{}
\usepackage[super]{cite}
\usepackage{xcolor}
\usepackage{amsmath}
\usepackage{amssymb}
\usepackage[verbose,hypertexnames=false]{hyperref}
\usepackage{physics}

\hypersetup{colorlinks=false,allbordercolors=blue,pdfborderstyle={/S/U/W 1}}

\begin{document}

\markboth{Leandro Morais et al.}
{Distinguishing Ordered Phases using Machine Learning and Classical Shadows}


\title{Distinguishing Ordered Phases using Machine Learning and Classical Shadows}

\author{Leandro Morais}

\address{{Instituto de Física de São Carlos, Universidade de São Paulo, CP 369, 13560-970 São Carlos, SP, Brazil. International Institute of Physics, Federal University of Rio Grande do Norte, 59078-970 Natal, Brazil}\\
leandro.silva@ifsc.usp.br}

\author{Tiago Pernambuco}
\address{Departamento de F\'isica Te\'orica e Experimental, Universidade Federal do Rio Grande do Norte, 59078-970 Natal-RN, Brazil}
\author{Rodrigo G. Pereira}
\address{International Institute of Physics, Federal University of Rio Grande do Norte, 59078-970 Natal, Brazil. Departamento de F\'isica Te\'orica e Experimental, Universidade Federal do Rio Grande do Norte, 59078-970 Natal-RN, Brazil}
\address{Departamento de F\'isica Te\'orica e Experimental, Universidade Federal do Rio Grande do Norte, 59078-970 Natal-RN, Brazil}
\author{Askery Canabarro}
\address{Grupo de F\'isica da Mat\'eria Condensada, N\'ucleo de Ci\^encias Exatas - NCEx,
Campus Arapiraca, Universidade Federal de Alagoas, 57309-005 Arapiraca-AL, Brazil. Quantum Research Center, Technology Innovation Institute, Abu Dhabi, UAE}
\author{Diogo O. Soares-Pinto}
\address{Instituto de Física de São Carlos, Universidade de São Paulo, CP 369, 13560-970 São Carlos, SP, Brazil}
\author{Rafael Chaves}
\address{International Institute of Physics, Federal University of Rio Grande do Norte, 59078-970 Natal, Brazil. School of Science and Technology, Federal University of Rio Grande do Norte, 59078-970 Natal, Brazil}

\maketitle

\newpage
\begin{abstract}
Classifying phase transitions is a fundamental and complex challenge in condensed matter physics. This work proposes a framework for identifying quantum phase transitions by combining classical shadows with unsupervised machine learning. We use the axial next-nearest neighbor Ising model as our benchmark and extend the analysis to the Kitaev-Heisenberg model on a two-leg ladder. Even with few qubits, we can effectively distinguish between the different phases of the Hamiltonian models. {Furthermore, by relying on a restricted set of local observables, such as pairwise correlations and plaquette operators, the sample complexity of the classical shadows protocol scales logarithmically with the number of measured features. This makes our approach a scalable and efficient tool for studying phase transitions in larger many-body systems where classical verification becomes intractable.}
\end{abstract}


\ccode{PACS number: 07.05.Mh}

\section{\label{sec:level1}Introduction}

Machine learning (ML) techniques have recently attracted considerable attention for their applications in quantum many-body physics \cite{carleo2017solving,torlai2018neural, carleo2019machine,carrasquilla2020machine,huang2020predicting}. One particularly promising area is their capability to distinguish between different phases and identify phase transitions across a range of Hamiltonian models \cite{carrasquilla2017machine,PhysRevB.99.121104, PhysRevLett.125.225701,broecker2017machine,rodriguez2019identifying,dong2019machine,wang2016discovering,che2020topological,kottmann2020unsupervised,kottmann2021unsupervised,uvarov2020machine,noronha2024predicting}. Traditional approaches, such as calculating order parameters and correlation functions, often face significant challenges in high-dimensional systems and may become impractical. Additionally, these methods are limited when dealing with systems exhibiting topological order, where phase transitions are not easily characterized by standard observables, cases in which alternative metrics, such as the Chebyshev distance, have shown greater effectiveness \cite{PhysRevB.102.134213}. By uncovering subtle patterns in quantum many-body data that indicate phase transitions, ML is proving instrumental in overcoming these limitations, offering new insights into complex quantum systems.

Despite significant advancements, ML methods remain heavily reliant on high-quality input data, and generating such data for quantum many-body systems is a formidable challenge due to the exponential growth of the Hilbert space with increasing system size. Classical computational techniques, such as quantum Monte Carlo simulations \cite{foulkes2001quantum} and matrix product state (MPS) tomography \cite{cramer2010efficient}, have achieved notable success but often struggle with the complexity and diversity of these systems. In turn, fault-tolerant quantum computers present a promising alternative by enabling more efficient simulations of quantum many-body systems, potentially improving data availability for ML models. These quantum devices leverage variational quantum methods to prepare ground states of Hamiltonians and quantum machine learning (QML) techniques to analyze information about these states. However, realizing this potential quantum advantage in fully variational approaches is hindered by significant theoretical and practical challenges, including barren plateaus and local minima\cite{cerezo2022challenges,bittel2021training}. To bypass these training complications, a pragmatic solution involves integrating quantum-generated data with classical algorithms. A pragmatic solution involves integrating quantum-generated data with classical algorithms, exploiting the strengths of classical machine learning to tackle problems beyond classical computational limits and, in some cases, rival QML models \cite{huang2021power}. Nonetheless, fully characterizing quantum systems remains constrained by the need for exponentially increasing measurements as the number of qubits grows.

Recent breakthroughs offer a promising approach to mitigate this measurement bottleneck. As demonstrated by Aaronson \cite{aaronson2018shadow}, much of the data traditionally deemed necessary for predicting certain properties of quantum systems is, in fact, redundant. This insight led to the development of the classical shadows protocol, a method that enables accurate predictions of both linear and nonlinear functions of quantum systems with substantially fewer measurements, all while maintaining controlled error margins \cite{huang2020predicting}. By significantly reducing the measurement overhead, classical shadows provide a powerful tool for overcoming the curse of dimensionality, making them invaluable for exploring quantum phases and facilitating ML applications in complex quantum systems.
 
In this work, we employ classical shadow techniques to efficiently estimate expectation values and classify the phases of two paradigmatic spin-$1/2$ chain models: the axial next-nearest neighbor Ising (ANNNI) model  \cite{selke1988annni} and the Kitaev-Heisenberg ladder  \cite{PhysRevB.99.195112}.  Both models are relevant for describing the magnetic properties of real materials and display rich phase diagrams with multiple ordered and disordered phases. The ANNNI model is significant as the simplest model in which different types of competing magnetic orders stem from the interplay between frustrated Ising interactions and quantum fluctuations induced by a transverse magnetic field. In turn, in the Kitaev-Heisenberg ladder, the regime of dominant bond-dependent Kitaev interactions harbors two Kitaev spin liquid (KSL) phases, which can be viewed as topological phases characterized by gapped Majorana fermion excitations \cite{Feng2007}, a nonlocal string order parameter, and double degeneracy of the entanglement spectrum \cite{PhysRevB.99.195112}. Distinguishing these KSL phases from the long-range ordered and disordered phases that emerge when Heisenberg interactions perturb the pure Kitaev model is a crucial challenge in the search for exotic magnetic states in low-dimensional systems \cite{Takagi2019}.

{Using randomized measurements, we efficiently generate data sets for both models. While we rely on two-point correlators to characterize the ANNNI model and the ordered phases of the Kitaev-Heisenberg ladder, we also estimate a six-spin plaquette operator, which is essential to accurately identify the Kitaev spin liquid phases. These combined features are subsequently analyzed using the KMeans algorithm \cite{lloyd1982least}, an unsupervised machine learning method for clustering and classification. Our results demonstrate that this approach is highly effective in distinguishing between different quantum phases, even for small system sizes.}

The paper is organized as follows. In Sec. \ref{Hamiltonian Models}, we introduce the Hamiltonian analyzed in this paper, where we discuss the  ANNNI  and Kitaev-Heisenberg models. Next, we provide an overview of classical shadows in Sec.  \ref{Classical Shadows} and explain the ML pipeline for the different models. Then, in Sec. \ref{Machi_learning_phase_transition}, we present the results of machine learning phase transitions, including the model outcomes. Finally, we conclude with a discussion of our findings in Sec. \ref{discussion}.

\section{Hamiltonian Models}
\label{Hamiltonian Models}
To demonstrate the effectiveness of classical shadows in generating data for phase distinction in many-body systems, we examine two Hamiltonian models with paradigmatic features: the  ANNNI model \cite{PhysRev.124.346, Villain81,Allen01,Rieger96,Guimaraes02,Beccaria07,Nagy11,cea2024exploringphasediagramquantum} and the Kitaev-Heisenberg model \cite{PhysRevB.99.195112, KITAEV20062, PhysRevLett.125.227202, KITAEV20032, PhysRevLett.105.027204, PhysRevLett.108.127203}. Details of these models are provided below.

\begin{figure}[t!]
    \centering
    \includegraphics[width=0.8\linewidth]{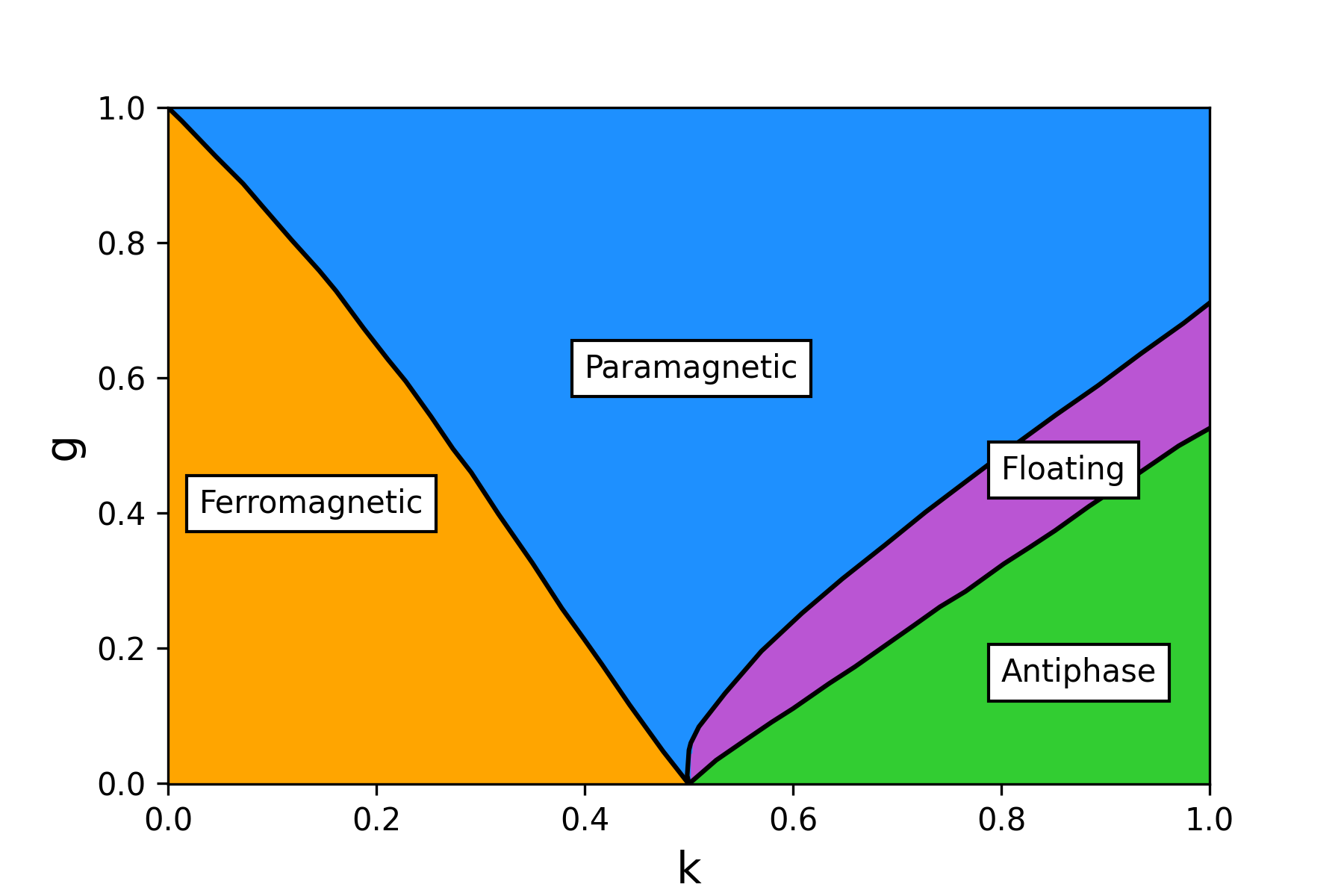}
    \caption{Phase diagram of the one-dimensional ANNNI model. The axes represent the frustration parameter $k = J_2/J_1$ (horizontal axis) and the relative transverse field strength $g = h/J_1$ (vertical axis). The diagram illustrates four distinct phases: Ferromagnetic (orange), Paramagnetic (blue), Antiphase (green), and the gapless Floating phase (purple), the sblack solid lines represents the transitions between these phases}
    \label{ANNNIDiagram}
\end{figure}

\subsection{THE ANNNI MODEL}
The ANNNI model was originally proposed to explain the magnetic ordering properties of rare-earth metals \cite{PhysRev.124.346}. 
 The  Hamiltonian for a one-dimensional chain is given by
\begin{equation}
\begin{aligned}
H = -J_1\sum_{ i} S^z_iS^z_{i+1} + J_2\sum_{  i} S^z_iS^z_{i+2} - h\sum_i S^x_i,
\end{aligned}
\label{ANNNI}
\end{equation}
where $S_i^\alpha$ with $\alpha=x,y,z$ are   spin-$\frac12$ operators at site $i$, $J_1>0$ describes a ferromagnetic Ising interaction between spins on nearest-neighbor (NN) sites, $J_2>0$ corresponds to a next-nearest-neighbor (NNN) antiferromagnetic interaction and the last term accounts for a transverse field acting on every lattice site. 

Despite its relative simplicity, the ANNNI model exhibits a complex phase diagram as the relative transverse field strength, $g = h/J_1$, and the ratio between NN and NNN interaction strengths, $k =J_2/J_1$, are varied.  This model is significant because it is the simplest system that combines the effects of quantum fluctuations---induced by the transverse field---and competing, frustrated exchange interactions. This interplay results in a rich ground-state phase diagram that has been studied extensively through analytical and numerical methods \cite{Villain81,Allen01,Rieger96,Guimaraes02,Beccaria07,Nagy11} . It consists of four distinct phases \cite{SELKE1988213,PhysRevB.100.045129}:

\begin{enumerate}
    \item For small $k$ and $g$, the system is in the ferromagnetic phase. In this phase, the system spontaneously breaks the $\mathbb Z_2$ symmetry $\sigma_j^z\mapsto -\sigma_j^z$. The two degenerate ground states are adiabatically connected to the product states in which the spins order themselves either as $\uparrow \uparrow \cdots \uparrow \uparrow$ or as $\downarrow \downarrow \cdots \downarrow \downarrow$;
    \item For large enough $g$, the transverse field term dominates and the system enters a paramagnetic disordered phase, with a unique ground state of spins aligned with the field;
    \item For large $k$, we have the so-called "antiphase", where the strong antiferromagnetic coupling between NNN spins causes the system to order with the pattern $\uparrow \uparrow \downarrow \downarrow \uparrow \uparrow...$, breaking the $\mathbb Z_2$ symmetry as well as translation invariance. The ground state is fourfold degenerate;
    \item For a small region of intermediate $k$ and $g$, the system is in a gapless  "floating phase" with power-law-decaying correlations. 
\end{enumerate}

The phase diagram of the ANNNI chain as a function of $k$ and $g$ is shown in Fig. \ref{ANNNIDiagram}.

The ANNNI model has a wide range of practical applications, including the description of rare-earth metals \cite{PhysRev.124.346,SELKE1988213}, the explanation of magnetic ordering in certain quasi-one-dimensional spin ladder materials \cite{Wen15}, the study of dynamical phase transitions \cite{Karrasch13}, and the exploration of interactions between Majorana edge modes in arrays of Kitaev chains \cite{Hassler12,Milsted15}. These diverse applications make it a paradigmatic model in many-body physics.

Building on the results in \cite{canabarro2019unveiling, ferreira2024detecting}, we will use the two-point correlation functions between all spins in the chain as input for the ML model. Unlike \cite{canabarro2019unveiling, ferreira2024detecting}, which relied on the diagonalization of the Hamiltonian, we will use classical shadows to estimate these expectation values in a much more efficient manner.

\begin{figure}[t!]
    \centering
    \includegraphics[width=1.0\linewidth]{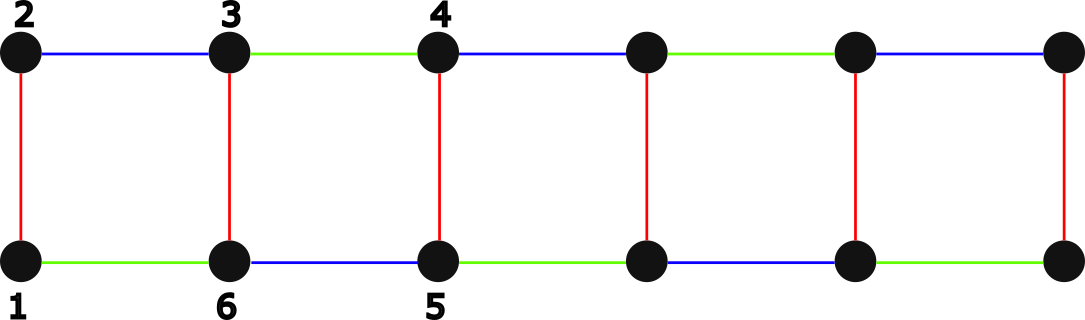}
    \caption{The Kitaev-Heisenberg ladder. Blue, green and red lines indicate $\gamma =$ $x$, $y$ and $z$ couplings respectively. The numbered sites on the ladder define the unit cell on which the plaquette operator in Eq. (\ref{Plaq}) is to be computed.}
    \label{Ladder}
\end{figure}

\subsection{THE KITAEV-HEISENBERG MODEL\label{KHmodel}}

Quantum spin liquids are among the most fascinating topics in contemporary condensed matter physics \cite{Savary2017,Broholm2020}, particularly since the introduction of the exactly solvable Kitaev model on a honeycomb lattice \cite{KITAEV20062}, which hosts a phase known as the Kitaev spin liquid (KSL). These KSL phases are of great interest due to their unique properties: when subjected to a magnetic field, they exhibit non-abelian anyonic excitations in the form of Ising anyons \cite{KITAEV20062, PhysRevLett.125.227202}. These excitations are particularly noteworthy because, in theory, they could be utilized for fault-tolerant quantum computation \cite{KITAEV20032}. 

\begin{figure}[t!]
    \centering
    \includegraphics[width=0.5\linewidth]{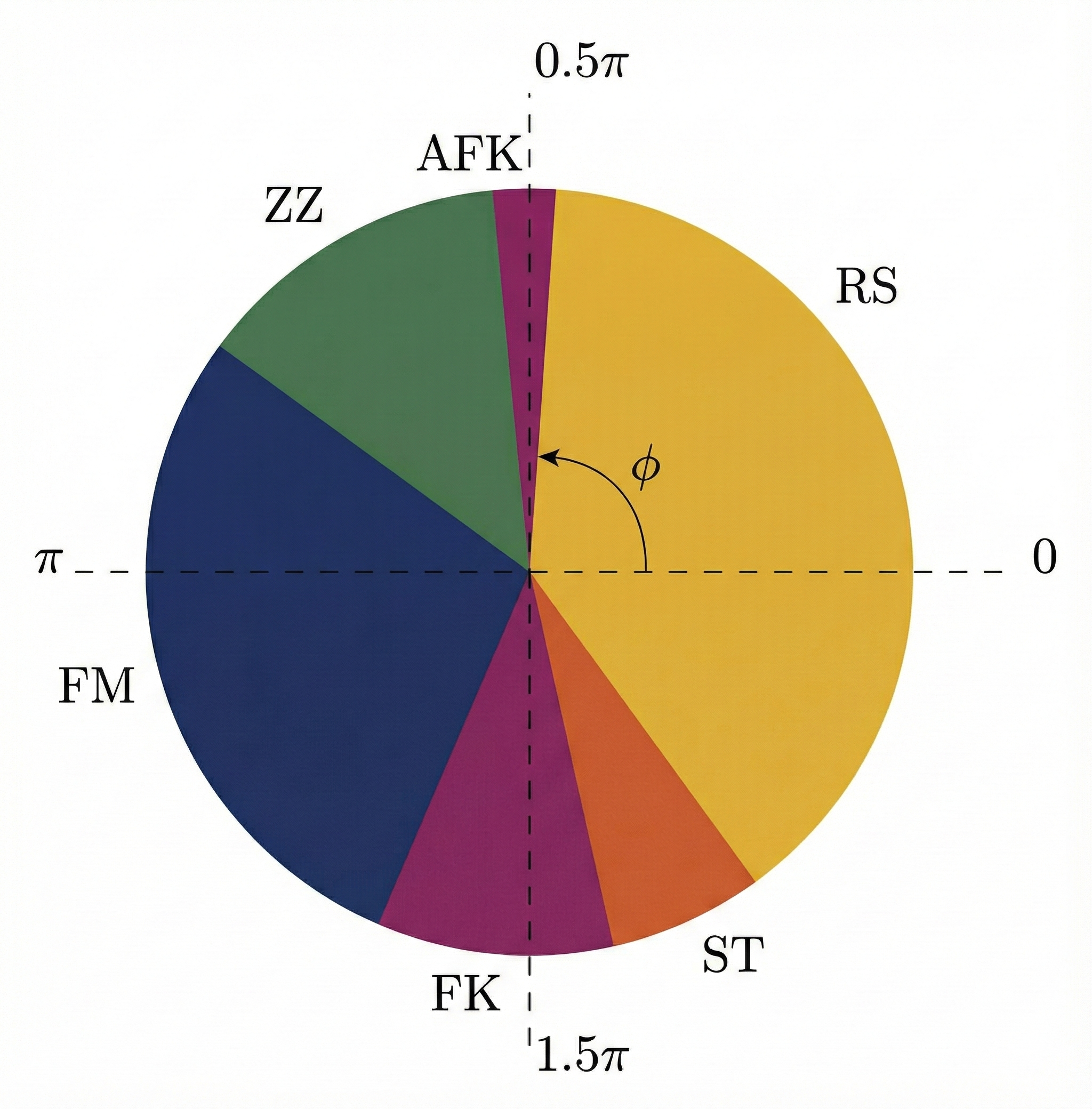}
    \caption{Phase diagram of the Kitaev-Heisenberg model as a function of the parameter $\phi$, which parametrizes the couplings as $K=\sin\phi$ and $J=\cos\phi$. The distinct magnetic phases are labeled by their acronyms: Ferromagnetic (FM), Zigzag (ZZ), Antiferromagnetic Kitaev spin liquid (AFK), Stripy (ST), Ferromagnetic Kitaev spin liquid (FK), and Rung-Singlet (RS).}
    \label{KH_Phases}
\end{figure}

Realistic models for candidate Kitaev materials, however, usually include perturbations beyond the exactly solvable model, for instance Heisenberg exchange interactions, which drive a wealth of magnetic orders in the vicinity of the KSL phase \cite{Rau2014,Rousochatzakis2024}. This extension leads to the Kitaev-Heisenberg  model, described by the  Hamiltonian 
\begin{equation}
\begin{aligned}
H = K \sum_{\gamma \in \langle i,j\rangle} S_i^{\gamma}S_j^{\gamma} + J \sum_{\langle i,j \rangle} \textbf{S}_i \cdot \textbf{S}_j,
\end{aligned}
\label{KH}
\end{equation}
where $\gamma =$ $x$, $y$ or $z$ depending on the specific bond and $\textbf{S}_i = (S_i^x, S_i^y, S_i^z)^T$, with the $S_i^{\alpha}$ being the spin operators acting on the $i$-th lattice site. The first term in the Hamiltonian represents the bond-dependent Kitaev interactions with coupling constant $K$ while the second one is the conventional isotropic Heisenberg interaction with coupling constant $J$. This model has been suggested to describe iridium oxides such as Li$_2$IrO$_3$ and Na$_2$IrO$_3$ \cite{PhysRevLett.105.027204, PhysRevLett.108.127203}.
The fact that the Kitaev-Heisenberg model can describe real materials with promising technological applications underscores the importance of understanding its behavior at large scales.

The Kitaev-Heisenberg model, like the original Kitaev model, was originally studied on a honeycomb lattice \cite{PhysRevLett.105.027204}. More recently, however, it has been shown that the Kitaev-Heisenberg model on a two-leg ladder lattice presents a phase diagram that is very similar to its honeycomb counterpart \cite{PhysRevB.99.195112, PhysRevB.99.224418}.

Here we will focus on the $S=\frac{1}{2}$  Kitaev-Heisenberg model on the two-legged ladder depicted in Fig. \ref{Ladder}. In this case, the Hamiltonian in Eq. \eqref{KH} can be decomposed as  \cite{PhysRevB.99.224418}
\begin{equation}
\begin{aligned}
H &= K \sum_{i=1}^{L/2} \left( S^x_{2i-1,1} S^x_{2i,1} + S^y_{2i,1} S^y_{2i+1,1} \right) + J \sum_{i=1}^{L} \mathbf{S}_{i,1} \cdot \mathbf{S}_{i+1,1} \\
&\quad + K \sum_{i=1}^{L/2} \left( S^x_{2i,2} S^x_{2i+1,2} + S^y_{2i-1,2} S^y_{2i,2} \right) + J \sum_{i=1}^{L} \mathbf{S}_{i,2} \cdot \mathbf{S}_{i+1,2} \\
&\quad + K \sum_{i=1}^{L} S^z_{i,1} S^z_{i,2} + J \sum_{i=1}^{L} \mathbf{S}_{i,1} \cdot \mathbf{S}_{i,2},
\end{aligned}
\end{equation}
where the $S_{i, j}^{\alpha}$ are the spin operators acting on the $i$-th spin on the $j$-th leg of the ladder and $L$ is the number of spins on each leg, making $N = 2L$ the total number of spins in the system.

\begin{figure}[t!]
    \centering
    \includegraphics[width=0.7\linewidth]{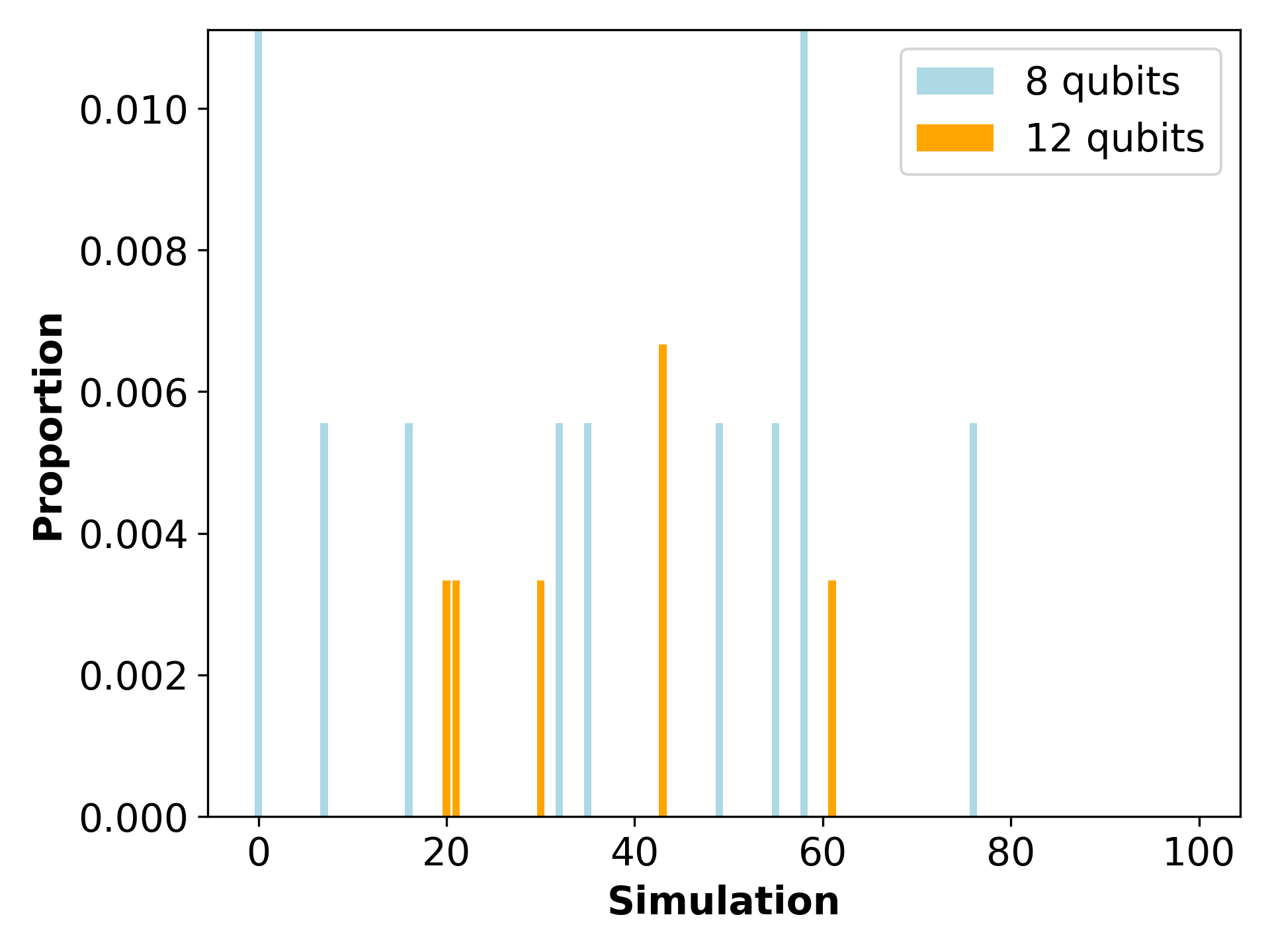} 
    \caption{ The plot displays the fraction of pairwise correlations ($\rho_{fail}$) that violated the error bound $\epsilon=0.1$ for each of the 100 simulations performed. Light blue and orange bars correspond to system sizes of 8 and 12 qubits, respectively. The low occurrence of errors confirms that the number of snapshots used \ref{snapshots} is sufficient for both system sizes.}
    \label{proportion} 
\end{figure}

The coupling constants for the Kitaev-Heisenberg model  can be parameterized as functions of a parameter $\phi$, with $K=\sin{\phi}$ and $J =\cos{\phi}$. Varying the value of $\phi$ reveals six different phases: four ordered phases and two spin liquid phases. They can be classified as follows \cite{PhysRevB.99.224418}:
\begin{enumerate}
    \item A rung-singlet (RS) phase for $-0.3\pi \leq \phi \leq 0.48\pi$, a trivial phase without magnetic order, where the unique ground state is adiabatically connected to a product of singlets on the rungs;
    \item An antiferromagnetic Kitaev spin liquid (AFK) phase for $0.48\pi < \phi < 0.53\pi$. This corresponds to a narrow region around the pure Kitaev limit at $\phi = \frac{\pi}{2}$, where $J = 0$ and $K = 1$;
    \item The zigzag (ZZ) phase (named after its analog phase in the honeycomb lattice) for $0.53\pi \leq \phi < 0.8\pi$. This phase has ferromagnetic order on each leg, but the  magnetization has opposite signs on different legs;
    \item A ferromagnetic (FM) phase for $0.8\pi \leq \phi < 1.37\pi$, in the region around $\phi = \pi$, where $J = -1$ and $K = 0$ ensure ferromagnetic Heisenberg interactions between the spins;
    \item A ferromagnetic Kitaev spin liquid (FK) phase for $1.37\pi \leq \phi \leq 1.57\phi$, the region around $\phi = \frac{3\pi}{2}$, where $J = 0$ and $K = -1$. Interestingly, this spin liquid phase is significantly wider than its antiferromagnetic counterpart;
    \item The stripy (ST) phase for $1.57\pi < \phi < 1.7\pi$, with long-range antiferromagnetic order along each leg but ferromagnetic correlations between spins on the same rung.
\end{enumerate}

The phases of the Kitaev-Heisenberg model described above are shown in Fig. \ref{KH_Phases}.

An interesting point to note about the AFK and FK spin liquid phases is that, although they do not present long-range order, they can be associated with near-unity expectation values for plaquette operators related to the flux of an emergent $\mathbb Z_2$ gauge field \cite{KITAEV20062}. On the two-leg ladder,  we consider the six-site plaquette operator  given by \cite{PhysRevB.99.224418}
\begin{equation}
    \mathcal{O}_{Plaquette} = { \sigma_1^x \sigma_2^y \sigma_3^z \sigma_4^x \sigma_5^y \sigma_6^z}, 
    \label{Plaq}
\end{equation}

for the sites chosen as shown in Fig. \ref{Ladder}. Strictly speaking, the expectation value of this plaquette operator does not qualify as an order parameter because it varies smoothly across the transitions out of the KSL phases. However, it can be calculated more easily than the string order parameter used in \cite{PhysRevB.99.195112}, with negligible finite-size effects, and drops rapidly in the long-range ordered phases \cite{PhysRevB.99.224418}.   

\begin{figure}[t!]
    \centering
    \includegraphics[width=0.8\linewidth]{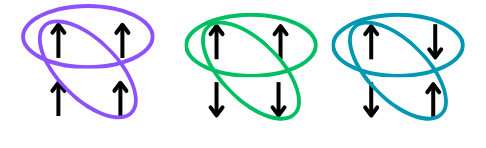}
    \caption{Conceptual illustration of pairwise correlations. The diagram shows how different spin configurations in a toy model lead to distinct patterns of pairwise correlations (e.g., between neighbors). These distinct patterns act as "fingerprints" for each phase, enabling the K-Means algorithm to group similar states into separate clusters.}
    \label{Correlations}
\end{figure}

The plaquette operators, as well as the two-point correlation functions between spins in the ladder, are the observables that we will estimate using classical shadows, which will then serve as the input data for identifying the different phases with a machine learning algorithm.

\section{An overview of the classical shadows method}
\label{Classical Shadows}

Classical shadows tomography is a method developed to efficiently estimate numerous properties of a quantum state, $\rho$, using a limited number of measurements. The procedure involves the repeated application of a randomized measurement protocol. Given an $N$-qubit quantum state $\rho$, a unitary operator $U$, sampled from a predefined ensemble, is applied to produce the rotated state $U\rho U^{\dagger}$. This state is subsequently measured in the computational basis, resulting in a collapse to the state $U^\dagger\ket{\hat{b}}\bra{\hat{b}}U$, where $\ket{\hat{b}}$ represents a classical bit-string. After the measurement, the inverse operation $U$ is applied to the post-measurement state. The resulting snapshot, $U^{\dagger}\ket{\hat{b}}\bra{\hat{b}}U$, contains valuable expectation information about the original state $\rho$, as described in the following equation:
\begin{align}
\label{expected}
\mathbb{E}[U^{\dagger}\ket{b}\bra{b}U] &= \mathbb{E}_{U \sim \mathcal{U}} \sum_{b \in \{0,1\}^n} \braket{b|U\rho U^{\dagger}|b} U^{\dagger}\ket{b}\bra{b}U \nonumber \\
&= \mathcal{M}(\rho) \text{.}
\end{align}
For any unitary ensemble $\mathcal{U}$, this process defines a quantum channel given by the mapping $\rho \mapsto \mathcal{M}(\rho)$. The classical shadow, denoted by $\hat{\rho}$, is the post-measurement state constructed such that, in expectation, it reproduces the original quantum state given by
\begin{equation}
\label{shadow_reconstruction}
  \hat{\rho} = \mathcal{M}^{-1}(U^\dagger \ket{b} \bra{b} U).
\end{equation}
In this work, we select as our unitary ensemble the set of unitaries formed by tensor products of $N$ single-qubit operators from the Clifford group, equipped with the uniform (Haar) measure over the unitary group $U(2^N)$. This specific choice of ensemble allows us to rewrite Equation~\ref{shadow_reconstruction} as
\begin{equation}
    \hat \rho  =  \bigotimes_{i = 1}^N \left(3U_i^\dagger |\hat b_i \rangle \langle \hat b_i|U_i - \mathbb{I}\right),
    \label{rhohat}
\end{equation}

where $U_i$ and $|\hat b_i \rangle$ are the Clifford unitary applied to the $i$-th qubit and the qubit's post-measurement state. For an arbitrary linear operator $O$ acting on $\rho$, we thus build an unbiased estimator $\hat o$ for the expectation value of $O$ written as\cite{huang2020predicting}
\begin{figure}[t!]
    \centering
    \includegraphics[width=0.7\linewidth]{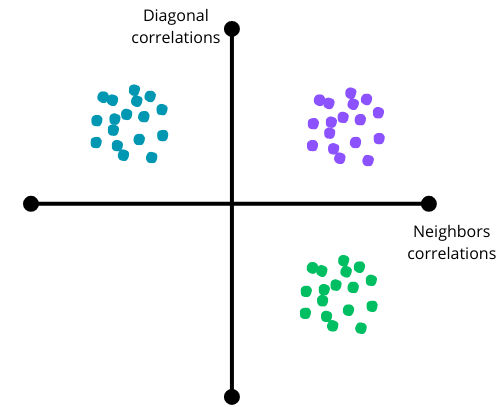}
    \caption{This scatter plot illustrates the effectiveness of the feature selection strategy described in Sec.~\ref{feature selections}, where the high-dimensional quantum state space is projected onto a reduced subspace defined by neighbor ($x$-axis) and diagonal ($y$-axis) pairwise correlations. 
This distribution aligns with the ``fingerprint'' concept illustrated in Fig.~\ref{Correlations}, demonstrating how the expectation values for ordered systems naturally organize into disjoint clusters, thereby facilitating phase distinction via unsupervised learning}
    \label{r2}
\end{figure}

\begin{equation}
    \hat o = \frac{1}{T} \sum_{m = 1}^T tr\left(O\hat \rho^{(m)}\right).
    \label{ohat}
\end{equation}

Above, $\hat \rho^{(m)}$ is the classical shadow associated with the $m$-th of a total of $T$ snapshots. The true power of equation \eqref{ohat} becomes clear once you consider a situation where you want to measure only the products of single qubit operators. Without loss of generality, we may consider an observable $O = \left[\bigotimes \limits^l_{i= 1} O_i\right] \otimes \mathbb I_{N-l}$. Taking this for the observable and \eqref{rhohat} for $\hat \rho^{(m)}$ in Eq. \eqref{ohat}, we have

\begin{equation}
    \hat o = \frac{1}{T}\sum \limits_{m = 1}^T\prod_{i = 1}^ltr(O_i \hat\rho_i^{(m)}),
\end{equation}

\noindent where $\hat \rho_i^{(m)} = \left(3U_i^\dagger |\hat b_i \rangle \langle \hat b_i|U_i - \mathbb{I}\right)$ for the $m$-th snapshot and we have used the property $tr\left(\bigotimes \limits^l_{i = 1} A_i\right) = \prod \limits_{i = 1}^l tr(A_i)$. Since the $O_i$ and $\hat \rho_i^{(m)}$ are $2 \times 2$ matrices, we see that it is entirely unnecessary to construct the entire $2^N \times 2^N$ density operator of the system to estimate a set of linear operators. Instead, we are left with the much easier task of evaluating a larger number of single-qubit expectation values and then multiplying the results.

Although simply taking the mean does lead to an estimator for the expectation value of $O$, it is typically more advantageous to partition the measurement data and employ a median of means to estimate observables \cite{huang2020predicting}. For linear observables that are the product of $N$ single-qubit operators, it was shown that the number of snapshots needed for estimating $M$ observables that act non-trivially on $l$ qubits with error less than or equal to $\epsilon$ is 
\begin{equation}
    T \propto \frac{\log(M)\cdot3^l}{\epsilon^2}.
    \label{snapshots}
\end{equation}

Equation \eqref{snapshots} can be turned into an upper bound for $T$ by considering the norms of the $M$ operators \cite{huang2020predicting}. However, known upper bounds are far from tight and tend to drastically overestimate the number of snapshots it takes to estimate an observable with a given precision \cite{huang2020predicting}. As such, in the present work, we fix the proportionality constant empirically based on numerical simulations in order to provide a tighter estimate of the actual number of snapshots necessary for the observables we intend to estimate here.

To perform this estimation, we set $\epsilon = 0.1$ and prepare the ANNNI and Kitaev-Heisenberg ground states for systems of 8 and 12 qubits. We then performed 100 trials estimations of the systems' pair correlation functions, keeping track of the number of errors above our $\epsilon$-threshold and how it evolves with system size, as shown in figure \ref{proportion}. Thus, we find that a good value for $T$ is

\begin{equation}
\label{snapshots}
    T = \frac{36 \log(M)}{\epsilon^2},
\end{equation}

\noindent where we take into account that $l = 2$ for pair correlations, which are the types of observables we will be working with throughout this work. Furthermore, it is important to highlight that while the advantage is marginal for small systems such as $N=12$, the protocol's logarithmic scaling ($\mathcal{O}(\log M)$) ensures scalability for $N \gg 12$, where direct measurement costs, which scale linearly with the number of observables, become prohibitive{Another tool we use in this work is the derandomization protocol of Huang et al\cite{huang2021efficient}. The derandomization protocol allows one to "take out" the randomness of classical shadows in the case where the measured observables are Pauli strings. This is accomplished by preselecting an advantageous sequence of measurements from the set of all random measurements corresponding to the chosen unitary ensemble \cite{huang2021efficient}.}

{Derandomizing the classical shadows protocol, which can be done in $B\times N$ iterations of the algorithm proposed in \cite{huang2021efficient} (for a $N$ qubit system with a measurement budget of $B$) tends to make it more efficient for the estimation of observables acting on many qubits at a time, providing a sort of "middle ground" between fully deterministic measurements of few global observables and random measurements of many local ones \cite{huang2021efficient}.}

\section{Performing the feature selection}
\label{feature selections}
Selecting the right data to collect is as crucial as obtaining them, as this decision directly influences the success of the analysis. While all information is encoded in the state vector describing the quantum system, extracting it through state tomography is exponentially expensive in terms of both measurements and data processing. This raises an important question: given the ground state of a Hamiltonian, how can we extract meaningful information about the quantum phase of this state without the need to fully reconstruct it? 

\begin{figure}[t]
\centering
\includegraphics[width=1\linewidth]{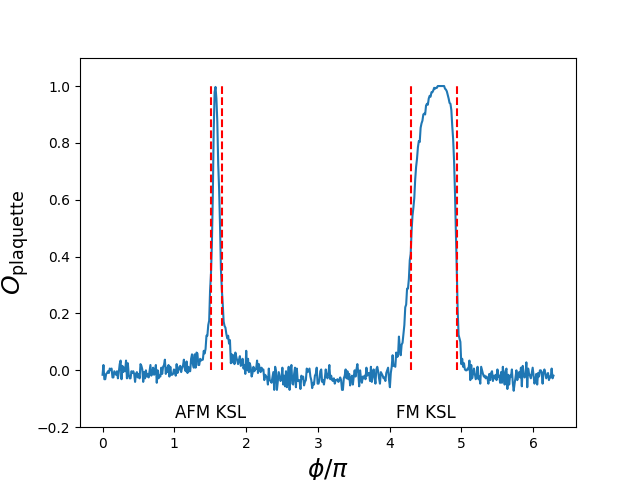} \caption{Detection of Spin Liquid phases. The plot displays the expectation value of the plaquette operator $\mathcal{O}_{Plaquette}$ (see Eq. \ref{Plaq}) as a function of the parameter $\phi/\pi$, calculated using derandomized classical shadows for a ladder of $N=12$ spins. The regions with distinct non-zero expectation values identify the Antiferromagnetic (AFK) and Ferromagnetic (FK) Kitaev Spin Liquid phases. The vertical dashed lines indicate the theoretical phase boundaries from. \cite{PhysRevB.99.224418}.} \label{PlaquetteOperator} 
\end{figure}

To illustrate this, consider the qualitative example shown in Fig. \ref{Correlations}. In this toy model, pairwise correlations between neighboring spins are sufficient to clearly indicate different phases. Specifically, for ordered systems where correlations are well defined, the expectation values form distinct clusters that can be easily identified. As a result, clustering algorithms such as K-means can efficiently group these values and thereby distinguish the corresponding phases, as visually demonstrated in Figure~\ref{r2}.

\begin{figure}[t!]
    \centering
    \includegraphics[width=1\linewidth]{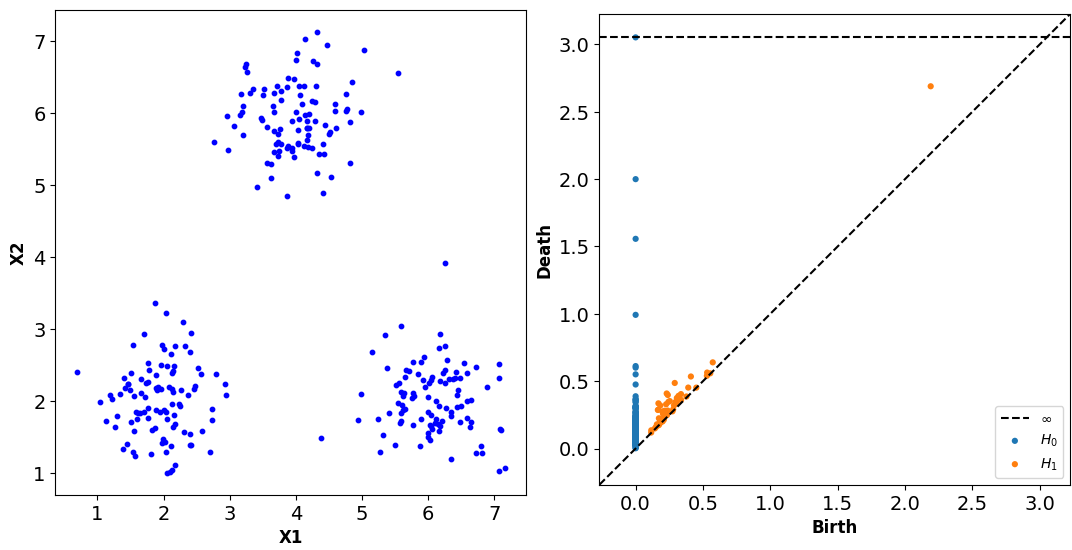}
    \caption{ Topological feature extraction. The left panel shows the spatial distribution of data points forming three distinct Gaussian clusters. The right panel displays the corresponding persistence diagram obtained via filtration. The blue points ($H_0$) located far from the diagonal represent connected components with high persistence, correctly identifying the existence of three stable clusters, while points near the diagonal represent topological noise}
    \label{TDA}
\end{figure}

A complementary approach to analyzing these clusters is provided by Topological Data Analysis (TDA), specifically through persistent homology. One of the main challenges when working with high-dimensional data is developing an intuitive understanding of its underlying structure. Persistent homology offers a qualitative method to capture the topological features of the data. In simple terms, persistent homology is constructed through a filtration process that generates \(k\)-simplices \cite{chazal2021introduction}. These \(k\)-simplices can be classified into (H$_0$) (connected components), (H$_1$) (structures with holes), (H$_2$) (cavities). As the filtration progresses, an increasing parameter \(\epsilon\) governs the appearance and disappearance of these \(k\)-simplices, enabling the construction of a birth-death diagram \cite{cao2024k}. These diagrams offer valuable insights into how homologies persist throughout the dataset, revealing its topological characteristics. For an illustration of these concepts, see Figure~\ref{TDA}.

In Figure~\ref{TDA}, three \(H_0\) structures persist throughout the filtration process. Intuitively, the points further from the diagonal represent longer intervals associated with topological features that persist for a significant duration during the filtration. These points are considered to have high persistence and are typically interpreted as important topological features, while points near the diagonal are regarded as noise in the data. If the clusters are too close together, persistent homology may struggle to effectively capture the clustering structure. Nevertheless, the information provided by the data can still be valuable from a qualitative perspective, as will be demonstrated in the case of the Kitaev-Heisenberg model. 
\subsection{Feature selection in the ANNNI model}
\label{Feature_selection_ANNNI}
Previous studies of the ANNNI model \cite{canabarro2019unveiling} have shown that the characterization of its main phases can be achieved taking as input the pairwise expectation values given by $\langle\lambda_0|S^{\alpha}_iS^{\alpha}_j|\lambda_0\rangle$, where $\alpha = {x, y, z}$, and $|\lambda_0 \rangle$ represents a ground state of the ANNNI model for given values of the parameters $g$ and $k$.
In the first study to classify ANNNI phases using machine learning \cite{canabarro2019unveiling}, all $N(N-1)$  pairwise correlators were used, resulting in a quadratic number of observables. However, as $N$ increases, this approach introducing redundant data and unnecessary computational complexitys. This challenge is especially relevant for algorithms implemented on quantum hardware, where limitations in qubit quality and quantity necessitate careful feature selection \cite{ferreira2024detecting}. Based on this reasoning, we have reduced the number of correlators to just those involving NN and NNN  sites, which, as will be shown, are already   enough to provide a precise characterization of the gapped  phases of the ANNNI model.

With this simplification, the number of  two-local observables to be estimated for each state is given by $N_{\mathcal{O}}$ $=  3(2N - 3)$, which scales linearly with the number of qubits. Consequently, the number of snapshots we need to perform using classical shadows is very efficient and given by
\begin{equation}
T = \frac{36\log(6N - 9)}{\epsilon^2}.
\label{Snapshot_annni}
\end{equation}

\begin{figure}[t!]
    \centering
    \includegraphics[width=0.8\linewidth]{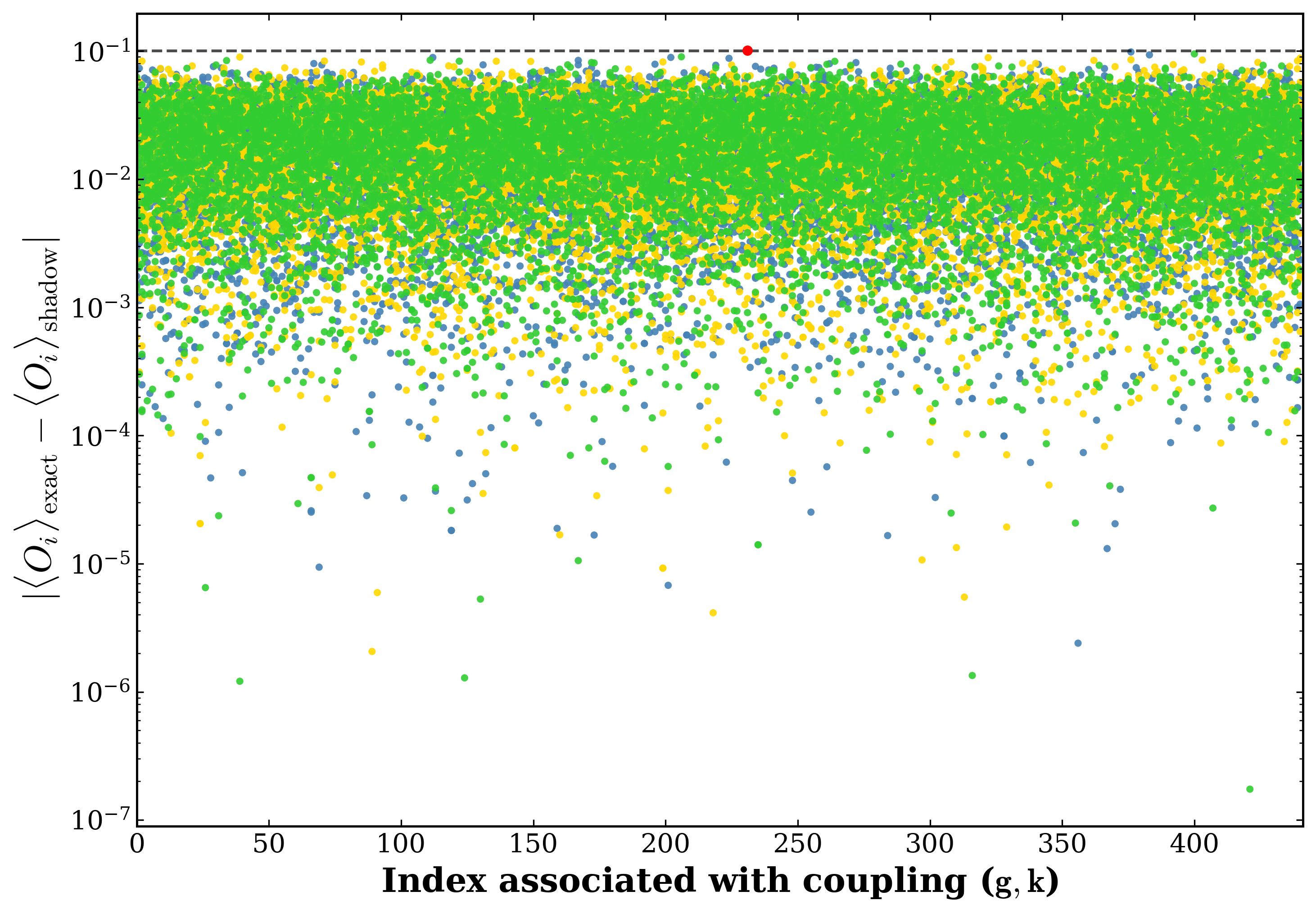}
    \caption{{Accuracy of correlation estimation. The scatter plot displays the absolute error of pairwise correlation estimates for the ANNNI model across various coupling configurations $(g, k)$ (horizontal axis). Colors indicate the correlation type: blue for X, yellow for Y, and green for Z. The horizontal dashed line marks the target error threshold $\epsilon=0.1$. The single red point represents an estimation outlier, which is expected within the statistical failure probability of the protocol}}
    \label{erro_figure_final}
\end{figure}

\subsection{Feature selection in the Kitaev-Heisenberg model}
\label{Feature-Kitaev}
For the analysis of the Kitaev-Heisenberg model, our initial approach is to examine the two-spin correlators. Although these correlators are typically essential for constructing phase diagrams with different types of magnetic order, they prove to be insufficient for detecting the spin liquid phases in the Kitaev-Heisenberg ladder. In this case, one could start with pair correlations and progressively extend the investigation to include higher-order terms. Phases of different natures may require correlations of a higher order; for instance, we expect quantum spin nematic phases, not observed in the models studied in this work, to be associated with correlators that involve spins on four sites \cite{shannon2006nematic}. 

However, for the Kitaev-Heisenberg model, we can take advantage of the fact that the expectation value of the plaquette operator, which can be understood as a six-spin correlator, had already been identified in the literature as a good indicator of phase transitions, as illustrated in Fig. \ref{PlaquetteOperator}. The plaquette operator acquires a nonzero expectation value of order unity inside the KSL phases. One can clearly identify a narrow region within the interval $\phi \in [0.48\pi, 0.53\pi]$, as well as a broader region within $\phi \in [1.37\pi, 1.57\pi]$, where the values exhibit noticeable deviations from zero. These deviations serve as clear indicators to distinguish between the spin liquid and ordered phases throughout the $\phi$ domain. Furthermore, the plaquette operator can be efficiently estimated through the derandomized classical shadows protocol \cite{huang2021efficient}, which offers a particularly suitable framework for this type of analysis.
\par 

\begin{figure}[t!] 
\centering \includegraphics[width=0.8\linewidth]{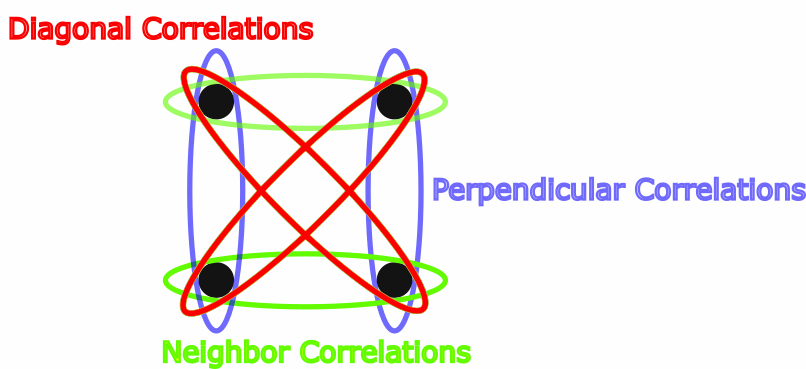} \caption{Schematic of the reduced feature set used for the Kitaev-Heisenberg model. The analysis is restricted to specific pairwise correlations: diagonal (red), perpendicular/rung (blue), and nearest-neighbor along legs (green).} \label{correlation quadrant} 
\end{figure}

Conversely, for the ordered phases, an initial analysis could consider all two-spin correlations. However, for the same reasons discussed in Section \ref{Feature_selection_ANNNI}, we restrict our analysis to the "quadrant correlations" shown in Fig. \ref{correlation quadrant}.   This set includes correlations between (NN) spins on the same rung and on each leg, as well as  (NNN) spins along the diagonals of the ladder. For a ladder of length $N/2$, the number of observables to be estimated for each value of $\phi$ is given by $N_{\mathcal{O}} = 3N - 6$,  requiring a total number of snapshots given by
\begin{equation} 
T = \frac{36\log(3N - 6)}{\epsilon^2}.
\label{snapshot_Kitaev} \end{equation}



 \begin{figure}[t!]
    \centering
        \includegraphics[width=1\linewidth]{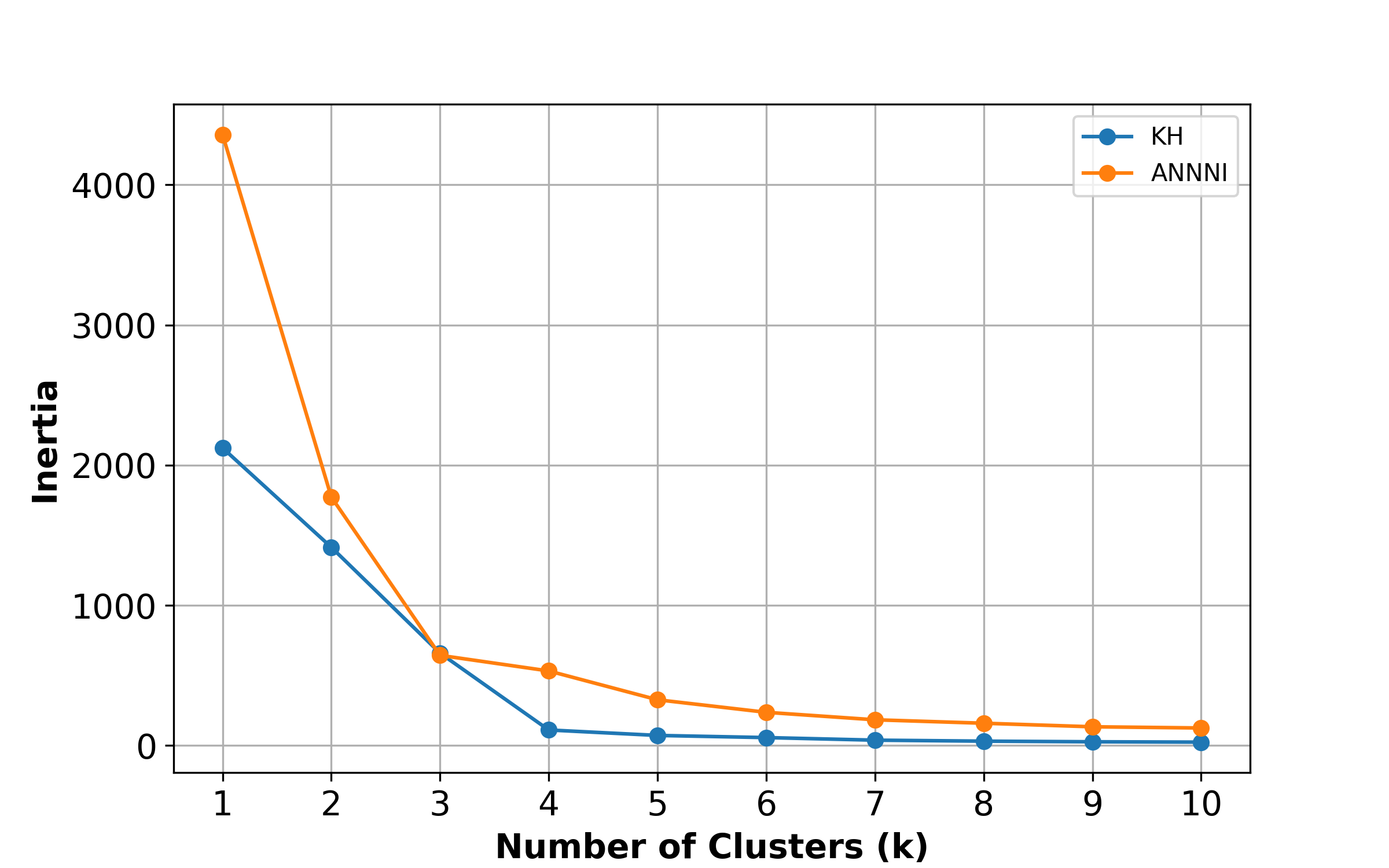}
\caption{{Elbow curves for optimal cluster determination. The inertia (sum of squared distances to centroids) is plotted against the number of clusters $k$ for the ANNNI (orange) and Kitaev-Heisenberg (blue) models. The ``elbow'' points suggest the optimal number of clusters for the K-Means algorithm.} }    \label{Elbow}
\end{figure}

\subsection{The K-Means Clustering Algorithm}
To classify the quantum phases based on the selected features, we employ the K-Means algorithm~\cite{lloyd1982least}, a widely used unsupervised machine learning technique. The algorithm partitions the dataset of $M$ observations into $k$ distinct clusters. It operates iteratively to minimize the inertia, defined as the sum of squared Euclidean distances between data points and their respective cluster centroids.

While other clustering algorithms, such as DBSCAN, have been successfully applied to identify quantum phases in similar contexts~\cite{canabarro2019unveiling}, we select K-Means—specifically Lloyd's algorithm—due to its predictable computational complexity. In this context, the complexity of K-Means is given by $\mathcal{O}(k \cdot M \cdot d \cdot i)$, where $k$ is the number of clusters, $M$ is the number of data points (ground states), $d$ is the number of features, and $i$ is the number of iterations until convergence. In our study, $k$, $M$, and the average $i$ are generally fixed or bounded. Therefore, the computational performance is primarily dictated by the number of features $d$.  

If one were to use all possible pairwise correlations, $d$ would scale quadratically with the system size, $\mathcal{O}(N^2)$. However, by employing our feature selection strategy (limiting inputs to neighbors and diagonals), we ensure that $d$ scales linearly with $N$. This choice guarantees that our protocol remains computationally efficient in the classical processing stage, complementing the logarithmic measurement efficiency provided by the Classical Shadows protocol.

\section{Machine learning phase transitions}
\label{Machi_learning_phase_transition}

\subsection{The ANNNI model}

As discussed in Sec. \ref{Hamiltonian Models}, the ANNNI model exhibits four distinct phases, characterized by three phase transitions. These transitions can be identified using the K-Means algorithm, fed with pairwise correlations of the ground states for $k,g$ couplings varying within the interval $[0,1]$. We used Eq.  (\ref{Snapshot_annni}) to compute the number of snapshots required to estimate the pairwise correlations with randomized measurements with a margin of error of 0.1 for each pair of 
$k,g$ couplings. The results are presented in Fig. \ref{erro_figure_final}. Only one of the estimations overshoots the bound. However, as discussed in \cite{huang2020predicting}, some probability of failure is expected. In this way, the classical shadow algorithm performs as expected, regardless of spin component or couplings.

With the data at hand, it is necessary to evaluate its structure to determine the optimal number of clusters for the K-Means algorithm. While several techniques are available for this evaluation, we opted to use the Elbow method \cite{syakur2018integration}, for which the "elbow" point indicates the best balance between variance explained and model complexity. The results are presented in Fig.  \ref{Elbow}, showing the inertia values for different numbers of clusters, which is crucial for deciding how many clusters to select before running the K-Means algorithm. It is evident that after 4 clusters, the inertia decreases more slowly compared to the interval between 0 and 4 clusters. Following the approach in \cite{canabarro2019unveiling}, we run the K-Means algorithm for 3 clusters, thereby leaving the floating phase aside.

\begin{figure}[t!]
    \centering
    \includegraphics[width=0.9\linewidth]{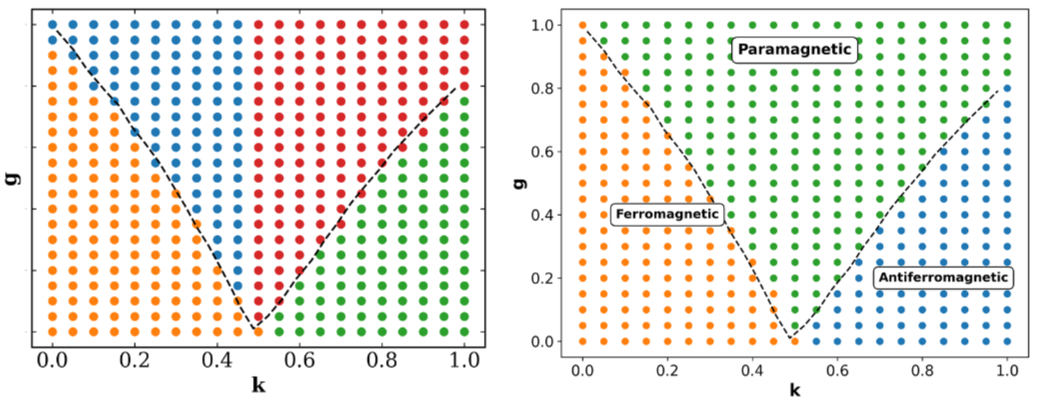}
     \caption{{Unsupervised phase classification of the ANNNI model for $N=12$ qubits. (Left) K-Means clustering results for $k=4$ clusters, demonstrating that the method cannot isolate the floating phase at this system size. (Right) Classification using $k=3$ clusters, which accurately recovers the Ferromagnetic, Paramagnetic, and Antiphase regions. The dashed black lines represent approximate theoretical phase boundaries}}
    \label{fasesquanticas}
\end{figure}

This approach allows us to obtain the phase diagram for the ANNNI model,  presented in Fig. \ref{fasesquanticas}, considering the Hamiltonian for 12 qubits. The result displays good agreement with the phase diagram obtained with much more numerical effort for larger system sizes \cite{Beccaria07}. We accurately determine the Ising transition between ordered (ferromagnetic) and disordered (paramagnetic) phases. The boundary between the paramagnetic phase and the antiphase falls in the region where we would expect to find the intervening gapless floating phase, which is not detected within this approach based on correlations for relatively small chains. { It is important to state this limitation explicitly: as evidenced by our negative results (for instance, at $k=4$), the floating phase cannot be isolated at $N=12$.}

The floating phase, as a critical phase, can be identified by the characteristic power-law decay of spin correlations at long distances. However, due to its presence in only a narrow region of the ANNNI model’s phase diagram, it has proven notoriously difficult to detect numerically, so much so that its very existence was long debated. Providing unambiguous numerical evidence for the floating phase demands simulations of much longer chains, typically involving hundreds of sites, which can be achieved using density matrix renormalization group (DMRG) methods \cite{Beccaria07}. An interesting direction for future research is to investigate whether, when combined with DMRG data for longer chains, our method could reliably detect the floating phase and other elusive critical phases in a variety of models.

\subsection{The Kitaev-Heisenberg model}

\begin{figure}[h]
    \centering
    \includegraphics[width=1\linewidth]{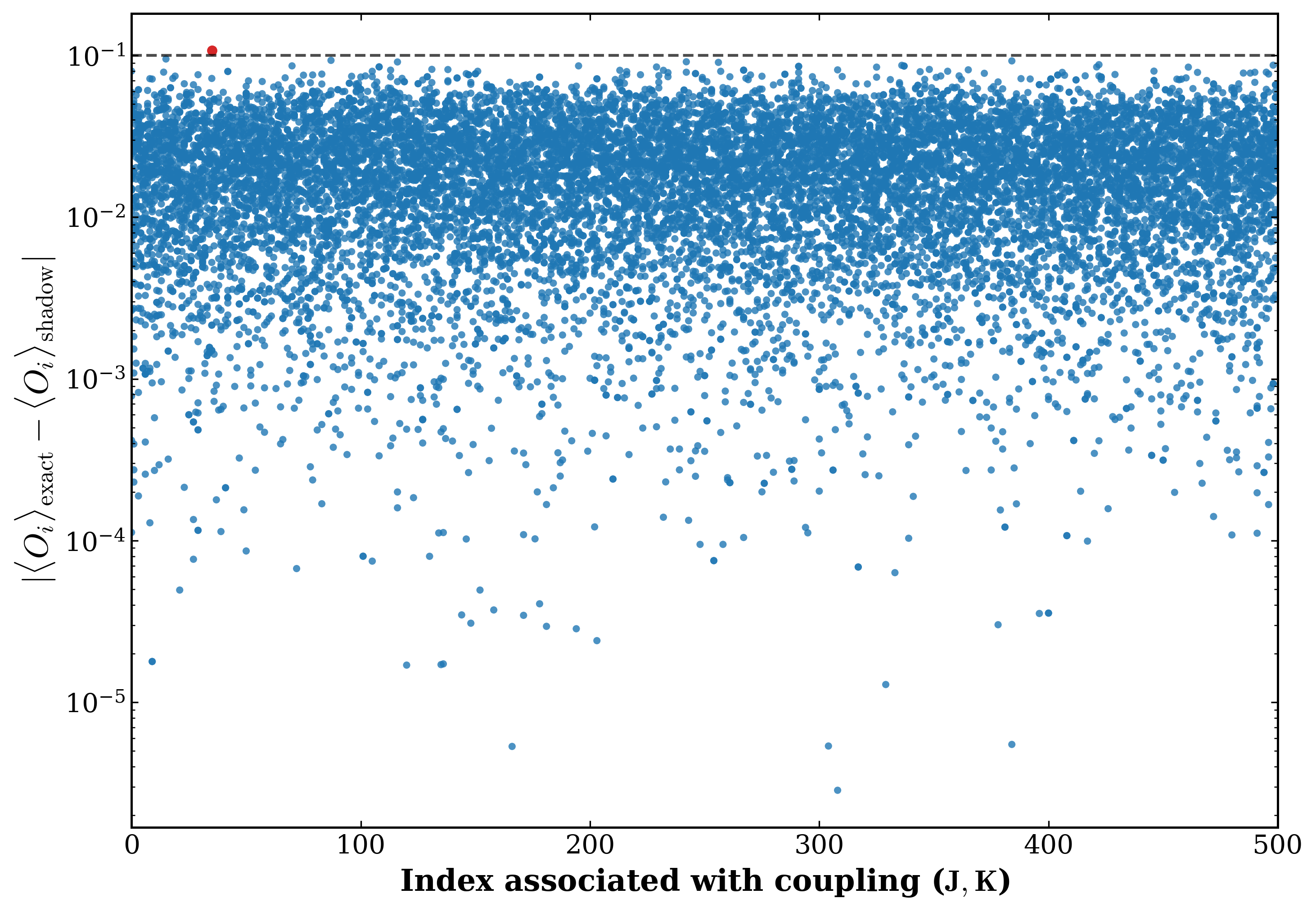}
    \caption{Validation of pairwise correlations estimation accuracy for the Kitaev-Heisenberg model. The scatter plot shows the absolute error $|\langle\mathcal{O}_i\rangle_{\text{exact}} - \langle\mathcal{O}_i\rangle_{\text{shadow}}|$ for pairwise correlations across the $(J, K)$ parameter space with a target threshold $\epsilon = 0.1$ (dashed line).}
    \label{ErrorBound_kitaev}
\end{figure}

Following the procedure outlined in Section \ref{Feature-Kitaev}, we estimated the pairwise correlations using classical shadows with a fixed error tolerance of $\epsilon = 0.1$. {For the derandomized classical shadow measurements used to estimate the plaquette operator, a budget of  $\mathcal O(10^3)$ measurements was sufficient to achieve that same error threshold} The results, presented in Fig. \ref{ErrorBound_kitaev}, show that for different couplings in the KH model, the pairwise correlations are accurately estimated, demonstrating the robustness of the classical shadows method. With the pairwise correlations accurately estimated, the subsequent step is to determine the number of clusters within the data from the ordered phases. As illustrated in Fig. \ref{Elbow}, the pairwise correlation data reveal four distinct clusters. This suggests the existence of four ordered phases, which is consistent with the discussion in Section \ref{Hamiltonian Models}. 
\par
\begin{figure}[h!]
    \centering
    \includegraphics[width=0.9\linewidth]{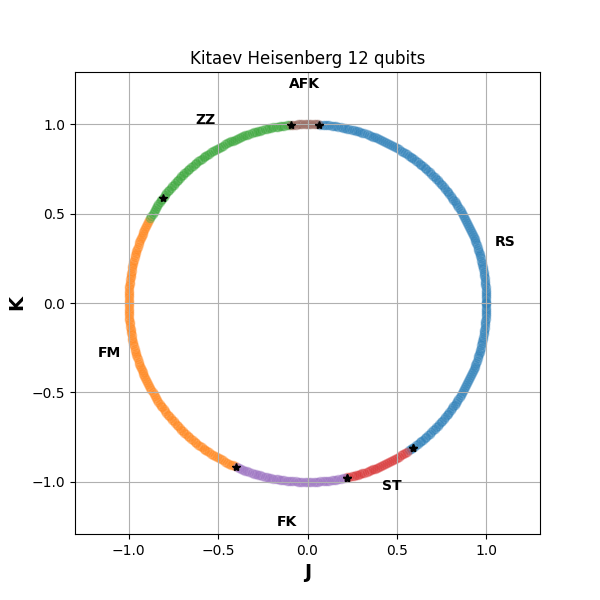}
    \caption{Reconstructed phase diagram of the Kitaev-Heisenberg ladder ($N=12$). The plot displays the phases in the $(J, K)$ parameter space identified by the hybrid pipeline using observables estimated via classical shadows. Colored regions correspond to: Rung-Singlet (orange), Zigzag (blue), Ferromagnetic (green), Stripy (red), and Kitaev Spin Liquids (purple/brown). Black stars mark theoretical transitions for comparison.}
    \label{FasesKitaev}
\end{figure}
\begin{figure}[t!]
    \centering
    \includegraphics[width=0.8\linewidth]{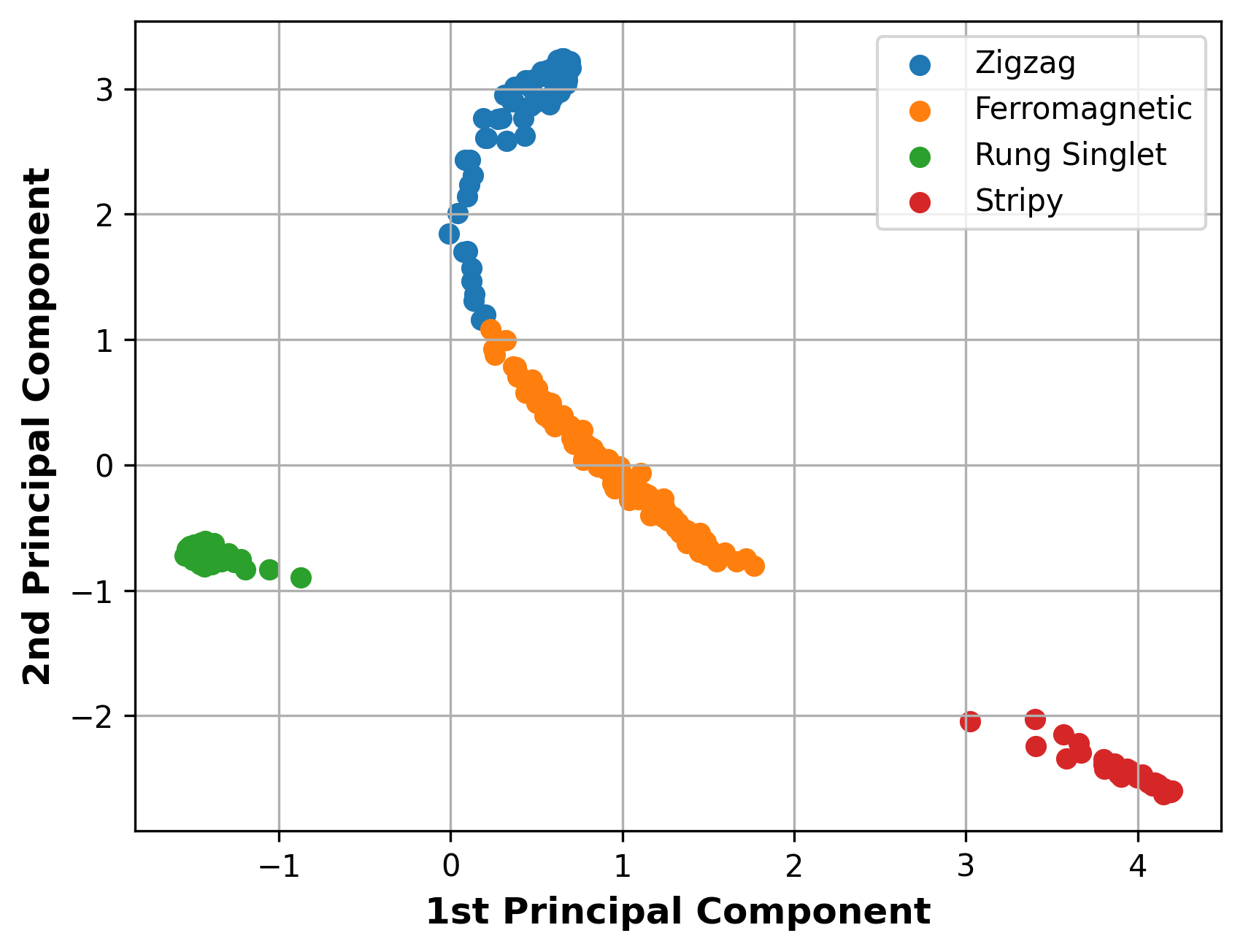}
    \caption{The scatter plot illustrates the projection of the high-dimensional correlation data onto the first two principal components for the 12-qubit Kitaev-Heisenberg model. Colors correspond to the distinct clusters identified by the K-Means algorithm: Zigzag (blue), Ferromagnetic (orange), Rung-Singlet (green), and Stripy (red). The close proximity between the Zigzag and Ferromagnetic clusters visually explains the reduced classification accuracy observed near the transition boundary between these phases.}
    \label{PCA}
\end{figure}

{To construct the final phase diagram shown in Fig. \ref{FasesKitaev}, we employ a step-by-step composite pipeline. First, we perform a pre-selection of the spin liquid sectors by evaluating the expectation value of the six-spin plaquette operator. Regions where this value exhibits significant non-zero deviations are isolated and identified as the Kitaev spin liquid phases. Second, for the remaining domain—where the plaquette operator vanishes and ordered phases reside—we apply the K-Means algorithm to the reduced set of pairwise correlations. This step successfully partitions the ordered sectors into four distinct clusters. By merging the results of these two steps, we obtain the complete phase diagram. It is important to emphasize that the black stars in the diagram represent benchmark transition points taken from the known phase diagram in previous literature \cite{agrapidis2019ground}, rather than boundaries independently inferred by our present method}. Notably, the transition between the zigzag and ferromagnetic phases appears less precise, while the transition from the stripy to the rung-singlet phase is much more accurate. This discrepancy occurs because with a limited number of qubits, the correlations for the zigzag and ferromagnetic phases become almost identical during the phase transition, making the two phases difficult to differentiate. To further explore this observation, we apply Principal Component Analysis (PCA) \cite{roweis1997algorithms} to the pairwise correlation data of the Kitaev-Heisenberg model. PCA is a statistical method that transforms the original dataset, which may include correlated variables, into a new set of uncorrelated variables called principal components. These components are ranked by the amount of variance they capture, with the first few components accounting for the most significant variations in the system. In our analysis, we projected the results onto the first and second principal components. The outcome is shown in Fig. \ref{PCA}.

\begin{figure}[t!]
    \centering
    \includegraphics[width=0.7\linewidth]{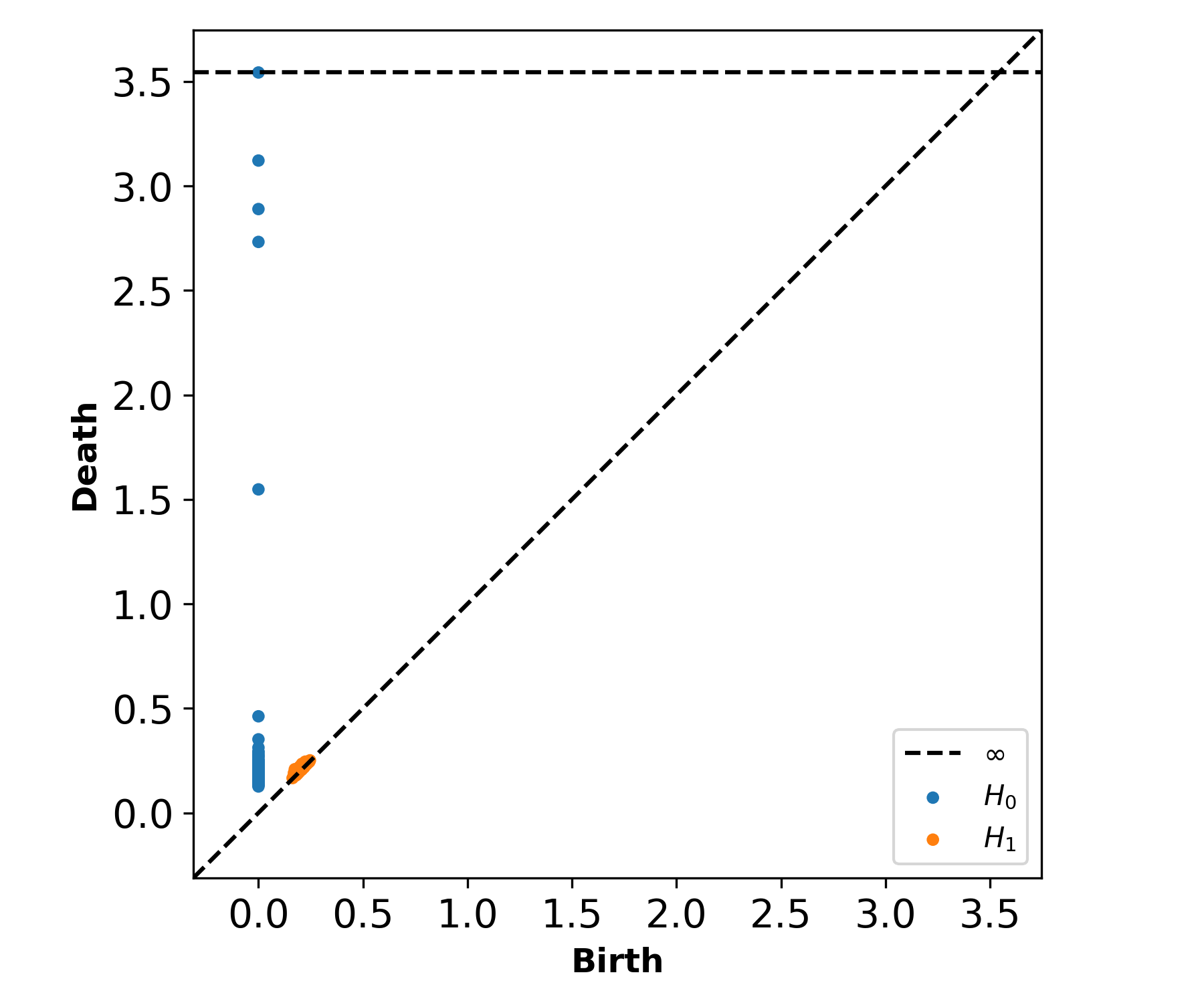}
    \caption{Persistence diagram for the Kitaev-Heisenberg model ($N=12$). The plot visualizes the lifespan of topological features during the filtration process. Blue points correspond to connected components ($H_0$), and orange points represent loops ($H_1$). The presence of four distinct $H_0$ features with high persistence (significant vertical distance from the diagonal) provides independent topological evidence for the existence of four clusters, corroborating the four ordered phases identified by the K-Means algorithm}
    \label{KHTDA}
\end{figure}

Except for the zigzag and ferromagnetic phases, the reduction of pairwise correlations of ordered phases into two principal components forms distinct groups, reflecting the different phases of the model. Furthermore, we can refer to the diagram of birth and death of persistence homologies in Fig. \ref{KHTDA} and identify which topological features are important, as discussed in \ref{feature selections}. For additional details, we recommend consulting \cite{chazal2021introduction}. As can be seen, we identify four structures (H0), indicating the presence of four clusters in these data points. The persistence diagram and PCA support the results obtained via the K-means algorithm, highlighting that the difficulty in pinpointing the exact transition between the zigzag and ferromagnetic phases arises from the weakly defined correlations of these phases when dealing with a small number of qubits. However, as the number of qubits increases, these correlations become more distinct. Consequently, we expect the K-means algorithm to identify the transitions with more precision.

\section{Discussion}
\label{discussion}

\begin{figure}[t!]
    \centering
    \includegraphics[width=0.8\linewidth]{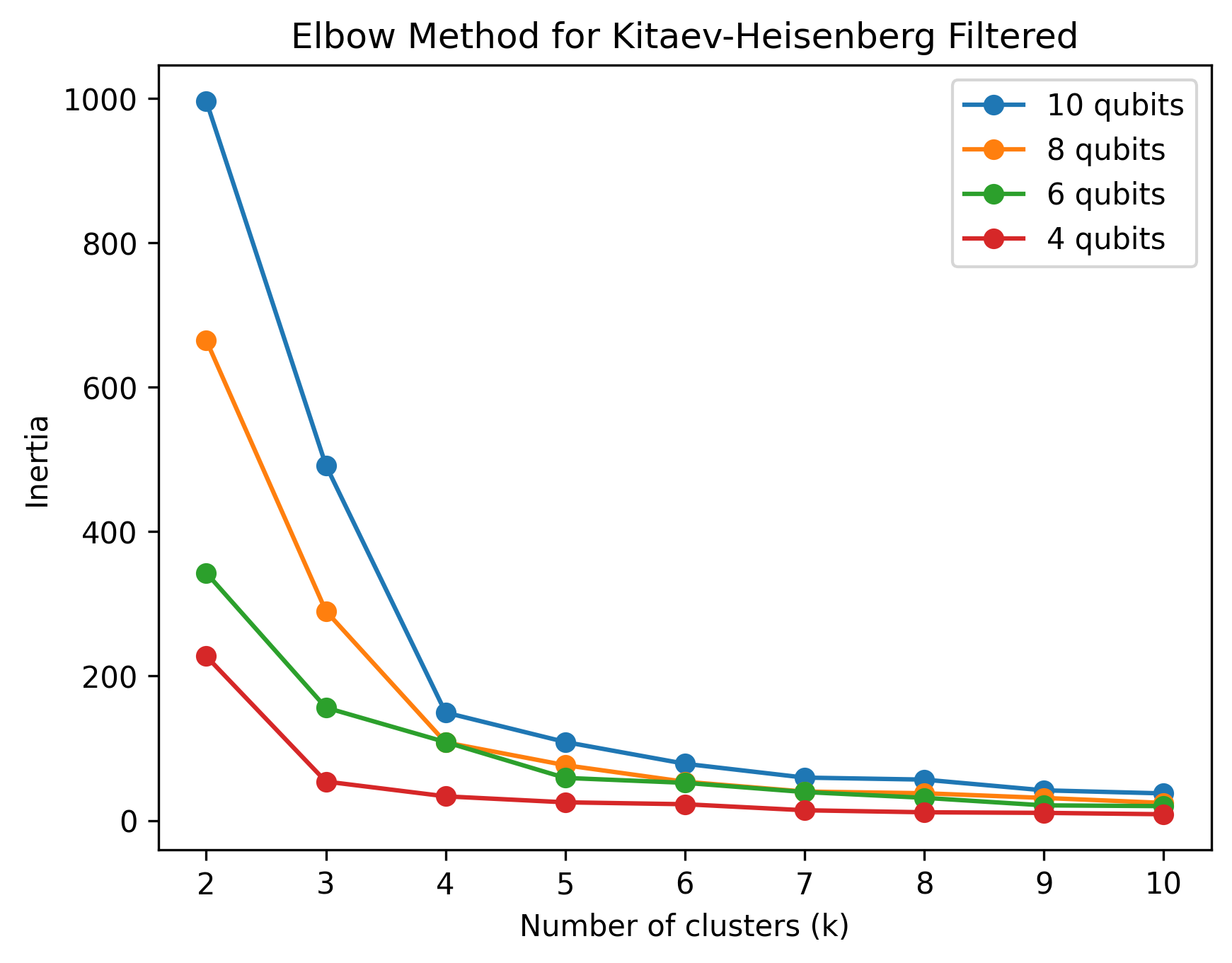}
    \caption{{Comparison between the filtered elbow curves for different system sizes in the Kitaev-Heisenberg model. By "filtered" we mean that the data for the spin liquid phases is excluded so that it is clearly visible that the regular ordered phases become more well-defined as the system size increases.}}
    \label{SizeComparison}
\end{figure}

This paper presents a quantum-classical pipeline for analyzing the phase diagram and phase transitions of quantum Hamiltonians. Using the ANNNI and Kitaev-Heisenberg models as benchmarks, we demonstrate how combining quantum computing with classical machine learning (ML) methods enables both qualitative and quantitative characterization of the complex phases in these Hamiltonians. {We begin by showing how observables can be efficiently estimated using the classical shadows protocol. Specifically, while two-point correlators are sufficient to characterize the ANNNI model and the ordered phases of the Kitaev-Heisenberg ladder, estimating a six-spin plaquette operator proved essential to successfully identify the Kitaev spin liquid phases}. These estimates are highly consistent, with errors well below ${\epsilon \leqslant 0.1}$ for both models, underscoring the effectiveness of the protocol in generating reliable data from quantum measurement that then can be fed into classical ML models. Next, we explore how K-means clustering, a classical ML algorithm, can process this data to accurately predict the three main phases (ferromagnetic, paramagnetic, and clustered antiphase/floating) of the ANNNI model and, with reasonable precision, recover the four ordered phases of the Kitaev-Heisenberg model (zig-zag, ferromagnetic, stripy, and rung-singlet). While the results for the transition between the zigzag and ferromagnetic phases are less well-defined for a 12-qubit system, it is expected that increasing the lattice size will improve the precision in witnessing these transitions. {This claim is supported by Fig. \ref{SizeComparison}, where it can be clearly seen that the number of phases identified by the elbow method increases with the number of qubits in the system, as finite-size effects become less relevant and the defining correlation patterns for the ordered phases become more well-defined. Particularly, we see that the elbow's position goes from $k = 3$ to $k = 4$ with increasing $N$, showing that two phases that couldn't be distinguished at too small a value of $N$ are sharply distinct for a larger one.}

Despite the significant challenges inherent in quantum machine learning pipelines, such as barren plateaus and local minima, our results reinforce that a hybrid quantum-classical approach can still be genuinely useful, even with a limited number of qubits and we hope to motivate further investigation in Hamiltonian models of increasing number of qubits and complexity.

\section{Acknowledgements} 
We thank Richard Kueng for the inspiring lectures ``The randomized measurement toolbox'' given in Natal and during the Paraty Quantum Information School that motivated the development of this work. We also thank the High Performance Computing Center (NPAD/UFRN) for providing computational resources. LM and TP acknowledge Marco Cerezo for fruitful conversations during the Second Quantum Computing School at ICTP-SAIFR. LM also thanks  Pedro Alcantra for the productive discussions at IFSC-USP. DOSP acknowledges the support by the Brazilian funding agencies CNPq (Grants No. 304891/2022-3 and No. 402074/2023-8), FAPESP (Grant No. 2017/03727-0 and No. 2023/03562-1) and the Brazilian National Institute of Science and Technology of Quantum Information (INCT/IQ). RC acknowledges the Simons Foundation (Grant Number 1023171, RC), the Brazilian National Council for Scientific and Technological Development (CNPq, Grant No.307295/2020-6 and 403181/2024-0) and the Otto Moensted Foundation visiting professorship. RGP acknowledges the Simons Foundation (Grant Number 1023171, RGP) and the Brazilian National Council for Scientific and Technological Development (CNPq, Grant No. 309569/2022-2 and 404274/2023-4). AC acknowledges partial financial support by the Alagoas State Research Agency (FAPEAL) (Grant No. APQ$2022021000153$), the CNPq (Grant No. 168785/2023-4) and by Centro de Competência do EDGE/UFAL.

\bibliographystyle{unsrt}
\bibliography{ws-sample}

\end{document}